\documentclass[twocolumn,aps,prd,epsf]{revtex4}
\usepackage[dvips]{graphicx}
\usepackage{epsfig}
\usepackage{amsmath}
\usepackage{color}

\newcommand{\be}{\begin{equation}}
\newcommand{\ee}{\end{equation}}
\newcommand{\bea}{\begin{eqnarray}}
\newcommand{\eea}{\end{eqnarray}}
\newcommand{\non}{\nonumber}
\newcommand{\bit}{\begin{itemize}}
\newcommand{\eit}{\end{itemize}}

\newcommand{\mbf}{\mathbf}
\newcommand{\q}{CG$\chi$QM }

\begin{document}

\title{Chiral symmetry crossover with a linear confining  \\ and temperature dependent quark-antiquark potential. }
\author{P. Bicudo}
\email{bicudo@ist.utl.pt}
\affiliation{CFTP, Dep. F\'{\i}sica, Instituto Superior T\'ecnico,
Av. Rovisco Pais, 1049-001 Lisboa, Portugal}

\begin{abstract}
Recently we developed a numerical technique to compute chiral symmetry breaking at $T=0$ with different current quark masses $m_0$,
including the current quark masses of the six standard flavours $u, \, d, \, s, \, c , \, b, \, t$.
We also fitted from Lattice QCD data the quark-antiquark string tension $\sigma$ dependence on temperature $T$.
We now utilize $\sigma(T)$ to further upgrade the chiral invariant and confinement quark model to finite temperatures $T \neq 0$.
We study the quark mass at finite $T$ and obtain the corresponding chiral crossover at $T=T_c$. 
The quark mass critical curve has a shape similar, but not identical, to the string tension critical curve.
In the case of the lightest quarks, the quark mass and the chiral condensate essentially
vanish at $T=T_c$, except for the small explicit chiral symmetry breaking $m_u$ and $m_d$.
\end{abstract}
\maketitle

\section{Introduction}

\subsection{Motivation.}

We address the chiral crossover in the quark model perspective. The chiral
crossover is a  cornerstone to understand the QCD phase
diagram 
\cite{CBM}, 
for finite $T$ and $\mu$. 
Notice that, after enormous numerical efforts, 
the analytic crossover nature of the finite-temperature QCD transition 
was finally determined by Y. Aoki {\em et al.}
\cite{Aoki:2006we,Aoki:2006br},  
utilizing Lattice QCD and physical quark, and reaching the continuum 
extrapolation with a finite volume analysis.
The QCD phase diagram is extensively researched at the experimental collaborations LHC, RHIC and FAIR.
This will help us to further understand how the universe evolves in the Big bang model of the cosmos.

\begin{figure}[t]
\center{
\includegraphics[width=0.8\columnwidth]{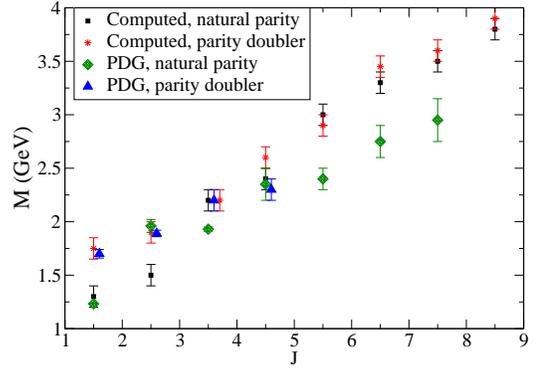} 
}
\caption{ First calculation of excited baryons with a chiral invariant quark model
\cite{Bicudo:2009cr}.}
\label{excitedBaryons}
\end{figure}

Here we utilize the linear confining potential for the
quark-antiquark interaction, in the  Coulomb Gauge Chiral Quark Model
(\q),
including both confinement and chiral symmetry
\cite{Bicudo:2010qp}. 
While this model, inspired from the framework of Coulomb gauge Hamiltonian 
formalism is not yet full QCD, it is presently the only model able 
to include explicitly both the quark-antiquark confining potential and 
the quark-antiquark vacuum condensation. 
The \q is for instance able to address excited
hadrons as in Fig. \ref{excitedBaryons},
and chiral symmetry at the same token, and it recently led us to suggest that
the infrared enhancement of the quark mass can be observed in the excited
baryon spectrum at CBELSA and at JLAB
\cite{Bicudo:2009cr,Bicudo:2009hm}.

Chiral symmetry breaking has been studied in detail 
with the \q in the chiral limit and at vanishing temperature.
It is quite well understood how, in the chiral limit of $m_0=0$, the quark develops a constituent
running mass $m(p)$ function of the momentum $p$.  $m(p)$ is a solution of the mass 
gap equation (equivalent to the Schwinger-Dyson equation) for the quark,
and this is important to understand the spontaneous breaking of chiral symmetry.
Although the \q is adequate to study the QCD phase diagram microscopically, the scientific community is only starting to explore 
\cite{Bicudo:1993yh,Battistel:2003gn,Antunes:2005hp,Glozman:2007tv,Guo:2009ma,Lo:2009ud,Kojo:2009ha,Nefediev:2009zzb},
the \q with a finite temperature
$T \neq 0$ and with a finite current quark mass $m_0\neq 0 $.
Notice that a finite quark mass is not only crucial for the study of the hadron spectra,
it is also relevant and for the study the QCD phase diagram.
In the phase diagram, a finite current quark mass $m_0$
affects the position of the critical point between the 
crossover at low chemical potential $\mu$ and the phase transition
at higher $\mu$. 
The present work, not only addresses the QCD phase diagram,
but it also  constitutes the first step to allow us in the future to
extend the computation of any hadron spectrum, say the Fig. \ref{excitedBaryons} computed in
reference \cite{Bicudo:2009cr}, to finite $T$ .

\begin{figure*}[t!]
\includegraphics[width=0.65\columnwidth]{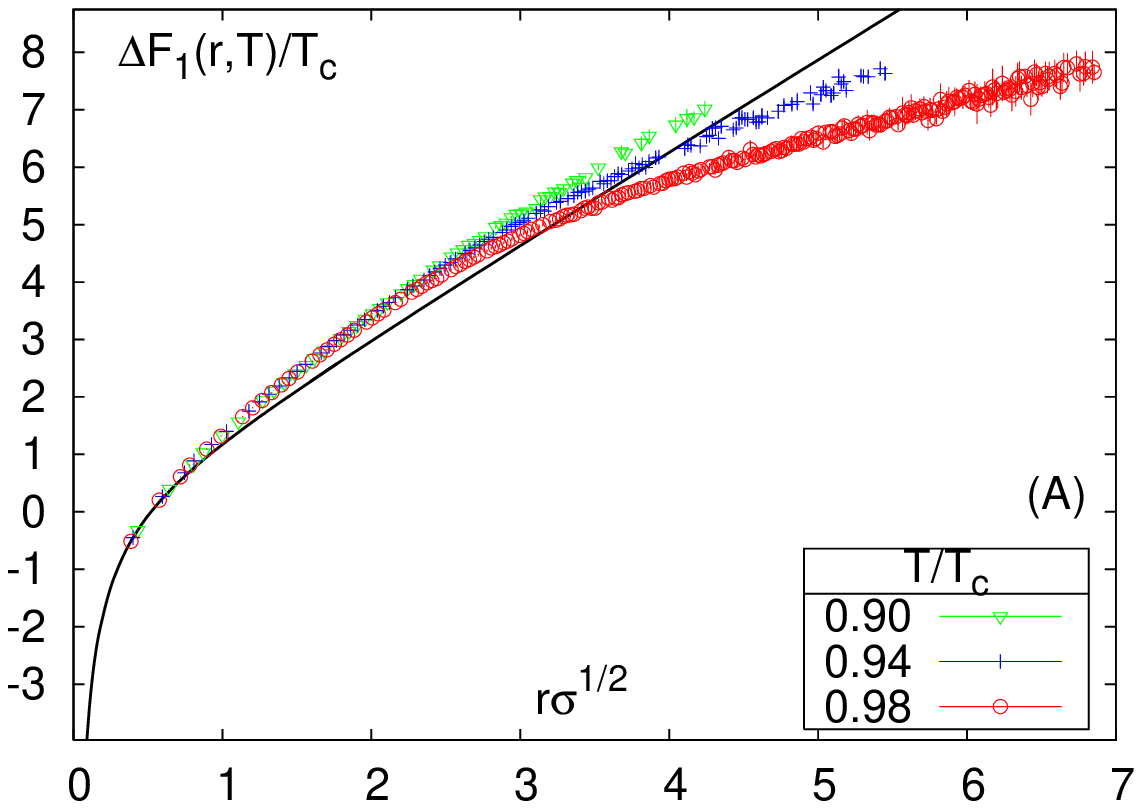}
\includegraphics[width=0.65\columnwidth]{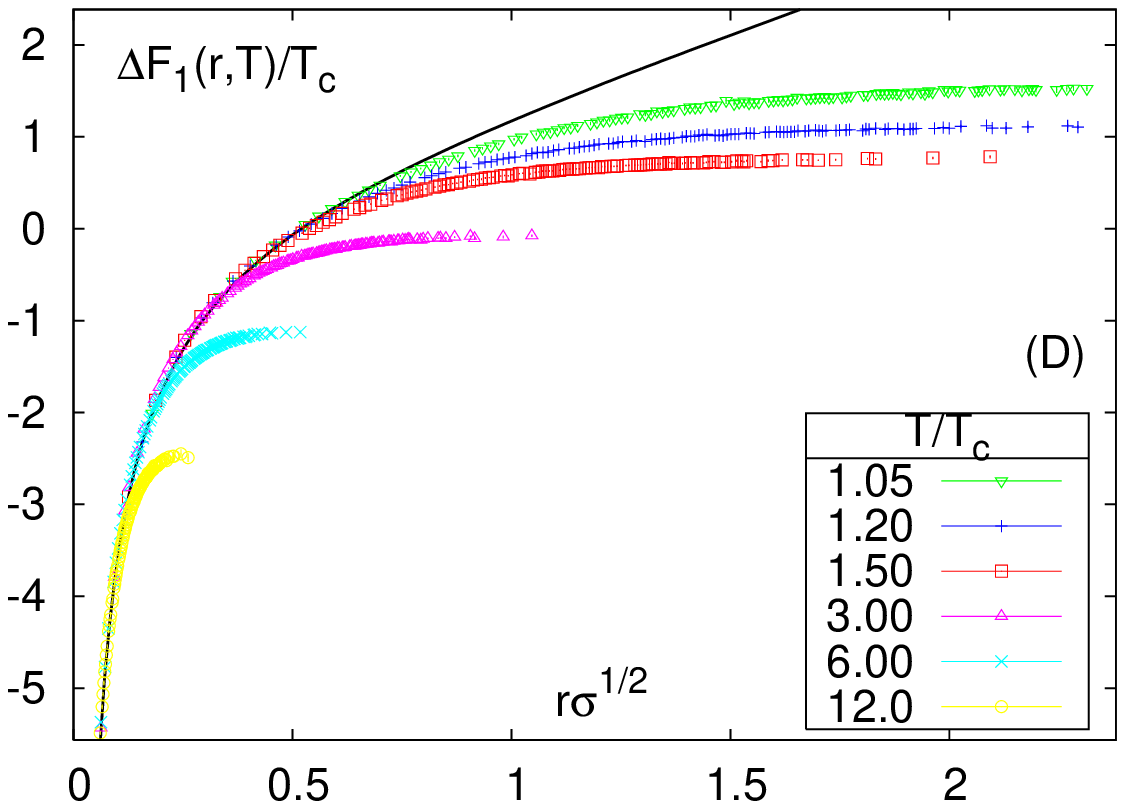}
\includegraphics[width=0.65\columnwidth]{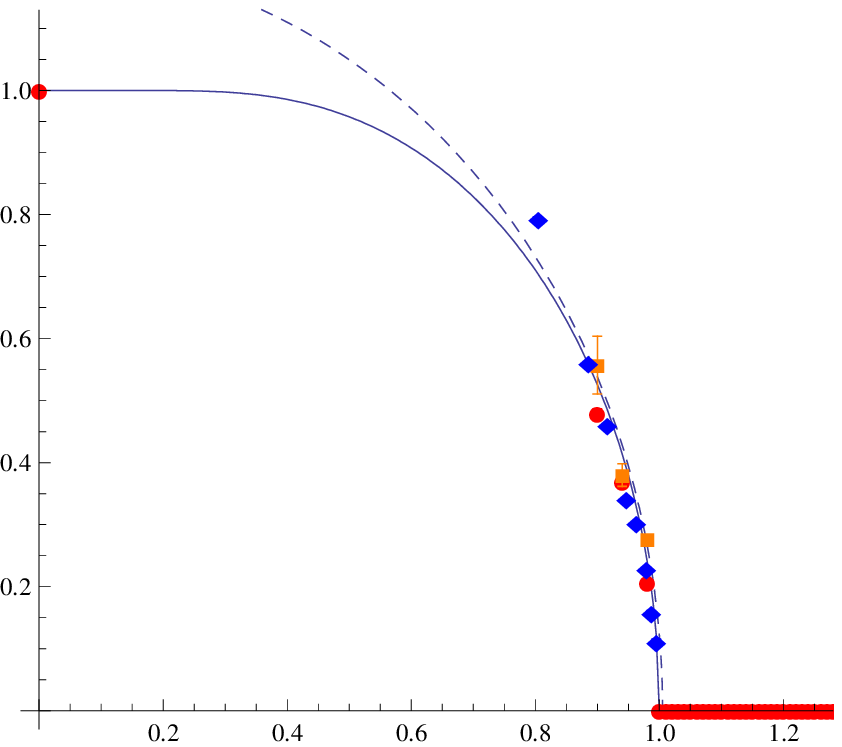}
\caption{\label{F1Kacz}
We compare the $T=0$  static quark-antiquark potential (solid line, left) to the  $T<T_c$  (left) and 
the $T>T_c$ (centre) 
Lattice QCD data for the free energy $F_1$, thanks to 
\cite{Doring:2007uh,Hubner:2007qh,Kaczmarek:2005ui,Kaczmarek:2005gi,Kaczmarek:2005zp}Olaf Kaczmarek et al.
We also compare (right) the magnetization curve (solid line, right) with the SU(3) 
string tension critical curve (bullets) computed in Ref. \cite{Kaczmarek:1999mm}.
We show (dashed line, right) the fit $1.21( 1- 0.990 (T/T_c)^2 )^{1/2}$ used in  Ref. \cite{Kaczmarek:1999mm}
to measure the finite string tension at $T_c$ as an evidence for a first order phase transition.
}
\end{figure*}

We now review recent advances, leading to a temperature dependent string tension, and to
a more efficient technique to solve the mass gap equation. These advances are applied in
Section II, where we derive the mass gap equation at finite Temperature.
In Section III we solve the finite $T$ mass gap equation, 
and compute the running mass $m_{_T}(p )$ to study the chiral crossover. 
In Section IV we discuss our results and conclude.

\subsection{The quark-antiquark potential at finite temperature $T\neq 0$.}

The most fundamental information for the quark-antiquark
interaction in QCD comes 
from the Wilson loop and from the Polyakov loops in Lattice QCD, 
providing the confining potential for a static quark-antiquark pair.
This potential is consistent with the funnel potential, also
utilized in the quark model to describe phenomenologically
the quark-antiquark sector of meson spectrum, in particular 
to describe the linear behaviour of mesonic Regge trajectories.
Notice that the short range Coulomb potential 
could also be included in the interaction, but 
here we ignore it since it only affects the quark 
mass through ultraviolet renormalization 
\cite{Bicudo:2008kc}, 
which is assumed to be already included in the 
current quark mass. Here we specialize in computing
different aspects of chiral symmetry breaking produced by
linear confinement $V=\sigma(T) \,r$. 

The finite temperature string tension was
computed for the first time by the Bielefeld Group for 
quenched SU(3) Lattice QCD 
\cite{Kaczmarek:1999mm}. 
The finite temperature static quark-antiquark free and
internal energies at some finite temperatures $T$ 
have also been computed in dynamical SU(3) Lattice QCD, by the Bielefeld for two light flavours $N_f=2$
\cite{Doring:2007uh,Hubner:2007qh,Kaczmarek:2005ui,Kaczmarek:2005gi,Kaczmarek:2005zp},
as depicted in Fig.  \ref{F1Kacz}.

Recently we have empirically shown the string tension at finite $T\neq0$ to be well fitted 
\cite{Bicudo:2010hg},
by a critical curve similar to the
spontaneous magnetization of a ferromagnet 
\cite{FeynmanLS},
{\em i. e.} solution of the
algebraic equation,
\be
{\sigma(T) \over \sigma (0) } = \tanh \left[  { T_c \over T}  {\sigma(T) \over \sigma (0) } \right] \ .
\label{eqformagnetization}
\ee
Although the empirical curve of Eq. (\ref{eqformagnetization})
corresponds to a second order transition (the solid line at the right of Figure \ref{F1Kacz}), 
and it is not a crossover  as in full QCD, or a first order transition as in pure gauge SU(3) QCD
(the dashed line at the right of Figure \ref{F1Kacz}),
the difference between these three scenarios is minute, and the validity of our fit is illustrated in 
Figure \ref{F1Kacz}.
Since the difference of these three scenarios has little effect in the numerics of the finite $T$ 
mass gap equation, the empirical curve of Eq. (\ref{eqformagnetization}) provides us with the necessary function 
$\sigma(T)$ to address confinement at finite temperature.

\subsection{The quark-antiquark potential, the mass gap equation 
and the Salpeter equation at $T=0$}

\begin{figure*}[t!]
\includegraphics[width=.7\columnwidth]{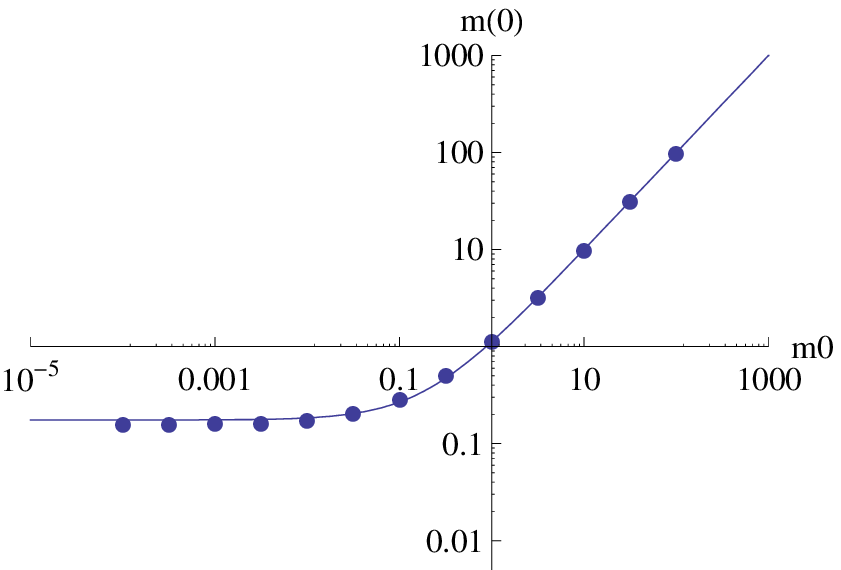}
\includegraphics[width=.7\columnwidth]{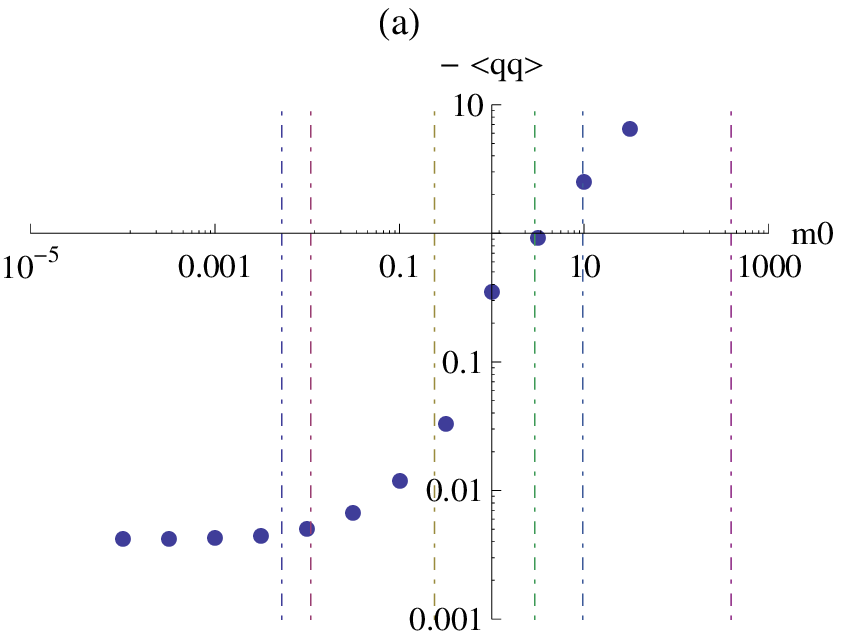}
\includegraphics[width=.55\columnwidth]{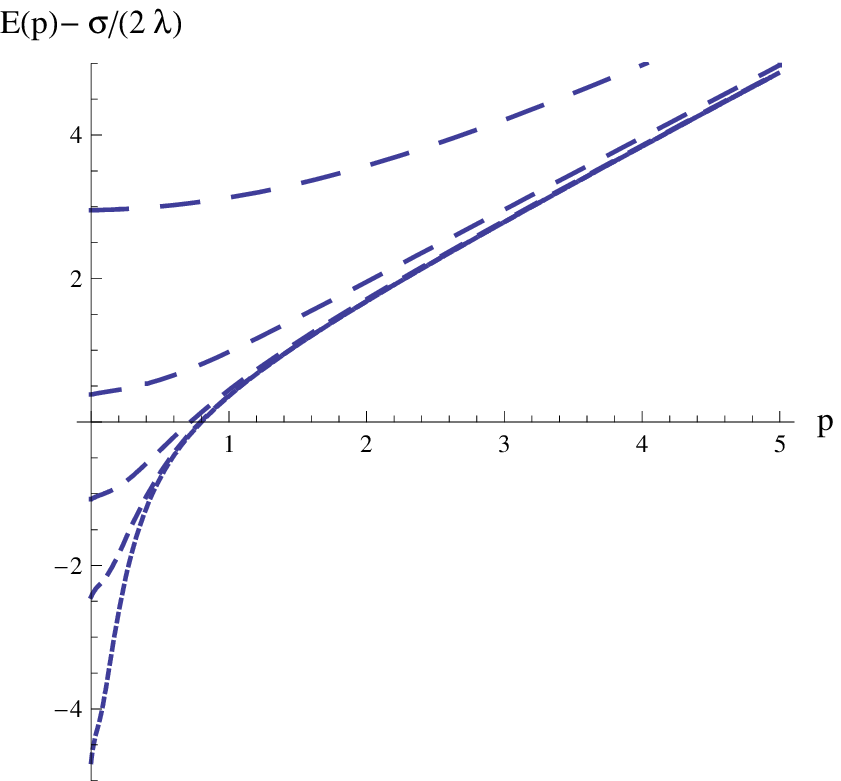}
\caption{ 
(left) We plot the solution of the mass gap equation $m(0)$ at $T=0$, 
for different values of the current quark mass $m_0$ in the case where the string tension
is $\sigma=1$, and we also show, with a solid line, the fit with a two-parameter irrational function.
(centre) We show a log log plot of minus the regularized quark condensate $- \langle \bar \psi \psi \rangle +  \langle \bar \psi \psi \rangle_0$
as a function of the current quark mass $m_0$, with vertical dot-dashed lines representing the
current masses of the quarks $u$, $d$, $s$, $c$, $b$, $t$.
(right) We represent the regularized quark dispersion relation $E(p) - {\sigma \over 2  \mu}$ 
with increasing number of dashes per curve 
for five different current quark masses $m_0$
with respective values $10^{-4}, 10^{-1}, 10^{-1/2}, 10^{0}, 10^{1/2}$.
The quark condensate has an inflection point 
for finite quark masses close to the strange quark mass, and 
for large masses it rises linearly with $m_0$.
All results are in dimensionless units of $\sigma=0.19$ GeV$^2=1$. 
\label{fitmassgap}}
\end{figure*}

To address the light quark sector it is not sufficient to know the static 
quark-antiquark potential, we also need to know what Dirac vertex to 
use in the quark-antiquark-interaction. This vertex is necessary to
study not only the meson spectrum but also the dynamical spontaneous
breaking of chiral symmetry.
To determine what vertex to use, we review how the quark-antiquark
potential can be approximately derived from QCD, in two different gauges.
In Coulomb gauge
\cite{TDLee}, 
\be
\mathbf \nabla \cdot \mathbf A(\mbf x, t)=0 
\ee
the interaction potential, 
as derived by Szczepaniak and Swanson
\cite{Szczepaniak:1995cw,Szczepaniak:1996gb},
is a density-density  interaction, with Dirac structure
$\gamma^0 \otimes \gamma^0$.
Another approximate path from QCD 
considers the modified coordinate gauge of Balitsky 
and in the interaction potential for the
quark sector,
retains the first cumulant order, of two gluons
\cite{Dosch:1987sk,Dosch:1988ha,Bicudo:1998bz}.
This again results in a simple density-density effective 
$\gamma^0 \otimes \gamma^0$ confining interaction. 
Assuming such a Dirac structure, our interaction potential for the quark sector is,
\begin{eqnarray}
V_I &=& \int\, d^3x \left[ \psi^{\dag}( x) \;(m_0\beta -i{\vec{\alpha}
\cdot \vec{\nabla}} )\;\psi( x)\;+
{ 1\over 2}  \int d^4y\, \
\right.
\nonumber \\
&&
\;\psi^{\dag}( \mbf x)
\lambda^a \psi ( \mbf x)  
{-3 \over 16}  \sigma(T) \, |\mbf x -\mbf y|
\;\psi^{\dag}( \mbf y)
\lambda^b 
 \psi( \mbf y)  
\label{hamilt}
\end{eqnarray}
where the Gell-Mann matrices are denoted $\lambda^a$,
and the density-density interaction includes just the linear confining potential
together with an infrared constant, which may be possibly divergent.
While the \q of Eq. (\ref{hamilt}) is not exactly equivalent to QCD,
we use it as our framework since it includes three crucial
aspects of non-pertubative QCD,  
a chiral invariant quark-antiquark interaction,
the cancellation of infrared divergences
\cite{Orsay1,Orsay2,Orsay3,Orsay4,Kalinowski,Lisbon1},
and  a quark-antiquark linear potential
\cite{linear1,linear2,linear3,Szczepaniak:1995cw,linear4,Wagenbrunn:2007ie}.
The mass gap equation and the energy of a quark are determined
from the Schwinger-Dyson equation at one loop order using the 
hamiltonian of Eq. (\ref{hamilt}). for a recent derivation
with all details see 
\cite{Bicudo:2010qp}.

In the limit of vanishing temperature $T=0$,
the interaction in the four momentum
of the potential and quark propagator term includes 
an integral in the energy 
\be
\int_{- \infty}^{+ \infty} {d \, p^0 \over 2 \pi}
{ i \over p^0 - E(\mbf p) + i \epsilon}  = + {1 \over 2 } \ ,
\label{minkowski}
\ee
which factorizes trivially from the vector $\mbf p$ momentum integral.
Using spherical coordinates, the angular integrals can be performed
analytically and finally only an integral in the modulus of the momentum 
remains to be computed numerically.
We arrive at the mass gap equation in two equivalent forms,
of a non-linear integral functional equation,
\bea
\label{massgabackbacktosincos}
0 &=& p S(p) - m_0 C(p) - { \sigma \over p^2}
\int_0^\infty {d k \over 2 \pi}  \, \bigl[
\\ \nonumber 
&&
I_A(k,p,\mu)\,  S(k) C(p) 
- I_B(k,p,\mu) \,  S(p) C(k) \bigr] \ ,
\eea
and of a minimum equation of the energy density $ {\cal E}$, 
\bea
\label{energybacktosincos}
&& {\cal E} = { -g \over 2 \pi}\int_0^\infty  {dp \over 2 \pi}
 \biggl[
2p^3 C(p) + 2 p^2 m_0 S(p) + \sigma \times
\\ \nonumber 
&&
\int_0^\infty {d k \over 2 \pi} 
 I_A(k,p,\mu) \,   S(k) S(p) 
+ I_B(k,p,\mu) \,   C(p) C(k) \biggr]  \ ,
\eea
where the functions $I_B$ and $I_A$ are angular integrals
of the Fourier transform of the potential.
In what concerns the one quark energy we get,
\bea
E(p) &= &  
p C(p) + m_0 S(p)  + {\sigma \over p^2}\, \int_0^\infty {d k \over 2 \pi}  I_A(k,p,\mu) \times
\label{regularized E}
\non
\\ 
&& 
\,    S(k) S(p)  +  I_B(k,p,\mu) \,   C(p) C(k)   \ .
\eea
We now extend this framework to finite $T$.

\section{Implementing finite $T$}

\subsection{The quark propagator at finite temperature $T\neq 0$.}

We now extend our equal-time density-density confinement to finite temperature.
A framework for this extension is the sum in Matsubara frequencies. 
With the equal time potential in our \q, the integration in the three-momentum $\mbf p$ and
in the energy $p^0$ are separable, and this is convenient to extend tour \q with a Matsubara.

\begin{figure}[t!]
\begin{center}
\includegraphics[width=0.9\columnwidth]{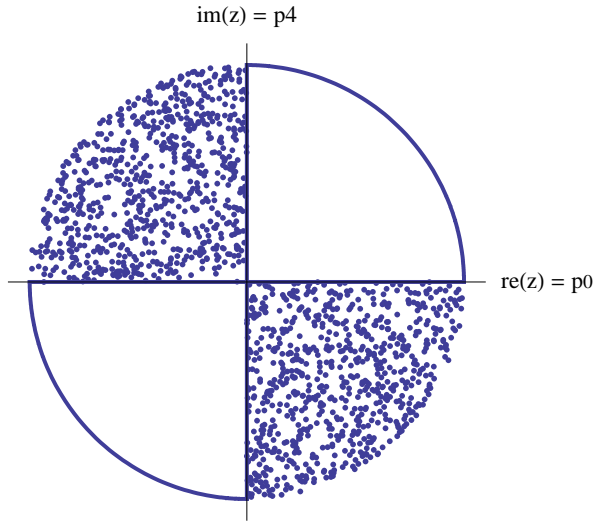}
\caption{Artist's view of the Wick rotation.
The Wick rotation from a real variable $z=p^0$ to a complex one $z=i \, p^4$,
can be performed without pole corrections, 
utilizing the depicted contour (solid in the figure), 
if there are only poles in the second and fourth Argand quadrants (dotted in the figure), 
and if the integrals in the circular paths of the first and third quadrants cancel in the limit of $|z| \to \infty$.
\label{Argand}}
\end{center}
\end{figure}

We first study the wick rotation as a link between Eq. (\ref{minkowski})  and the Matsubara sum.
In the $T=0$ mass gap equation or Schwinger Dyson equation, 
we have the Minkowski space integral in $p^0$ of the quark propagator pole of
Eq. (\ref{minkowski}) and this is equivalent to an integral 
in $p^4$ in Euclidian space after a Wick rotation in the Argand space,  
\be
\int_{- \infty}^{+ \infty} {i d \, p^4 \over 2 \pi}
{ i \over i p^4 - E(\mbf p) + i \epsilon}    = + {1 \over 2 } \ .
\label{euclides}
\ee 
This simple Wick rotation can be understood considering the closed contour  of Fig. \ref{Argand}
\bea
&& \left(  \int_{- \infty}^\infty + \int_\infty^{i \, \infty} + \int_{i \, \infty}^{-i \, \infty}+ \int_{-i \, \infty}^{- \infty} \right) 
d z f(z)
\\ \non
&=& 
\sum_{p_i  \in \mbox{ quadrant 1}} Re \, f(p_i) 
+ \sum_{p_j  \in \mbox{ quadrant 3}} Re \, f(p_j) \ .
\eea
In the case there are no poles in the first or third quadrant and the
circular paths cancel,  one just has to replace $p^0$ by $i \, p^4$ since in the real axis the path corresponds to $z=p^0$ and
in the imaginary axis the path corresponds to $z = i \, p^4$. 

Notice that the integral in Eq. (\ref{euclides}) is only identical to the one in Eq. (\ref{minkowski})  if the one quark
energy $E(p) >0 $. If $E(p) <0 $ the integral of Eq. (\ref{euclides}) changes sign, and this is consistent with the pole
moving from the fourth quadrant to the third quadrant of the Argand plane. While the integral in Eq. (\ref{minkowski}) 
is insensitive to this translation of the pole, the contour of Fig.  \ref{Argand} leads to a pole correction. For simplicity,
we choose to work with positive one quark energies only $E(p) >0 $.

In finite temperature $T$ and density $\rho$, the continuous euclidian space 
integration of eq. (\ref{euclides})
is extended to the sum in Matsubara frequencies,
\bea
\sum_{n=- \infty}^{+ \infty} i \, k T
{ i  \over i \, (2 n +1) \pi \, k T - \left[E(\mbf p) - \mu\right]} 
\non \\
= {1 \over 2} \tanh { E(\mbf p) - \mu \over 2 k T}
\label{matsubara}
\eea
where a chemical potential $\mu$ may be included, as a finite density when a Fermi sphere of quarks is present. 
It is clear that in the vanishing temperature and density $kT  << E $ limit one gets back the initial 
Euclidean space integral of eq. (\ref{euclides}), when the Matsubara sum approaches 
the continuum integration with $ k T \to {d P^4 \over 2 \pi}$.

\begin{figure}[t]
\hspace{0cm}
\begin{center}
\includegraphics[width=.9\columnwidth]{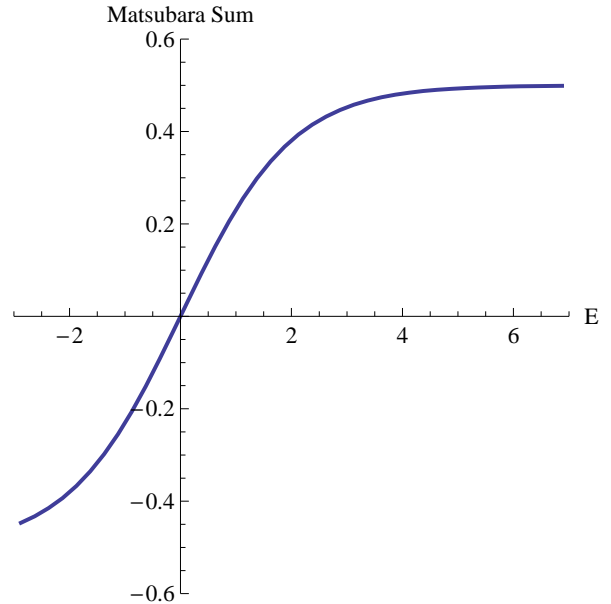}
\end{center}
\caption{The sum in Matsubara frequencies of the quark propagator as a function of $ E - \mu \over KT $,
including also energies below the fermi surface. The $T \to 0$ or $E - \mu \to \infty$ limit of the sum is $1 /2 $, as in Eq. (\ref{minkowski})
or in Eq. (\ref{euclides}).
\label{matsubaras}}
\end{figure}

In what concerns the analytical calculations leading to the results in Eqs.
(\ref{minkowski}) , (\ref{euclides}) and (\ref{matsubara}), we used the even
integral or summation to change the variable to a square, and we respectively
used the residue theorem, real and indefinite integrals, and the analytical series summation,
\be
\sum_{n=0}^\infty { y \over y^2 + (n+1 / 2)^2 }= { \pi \over 2 } \tanh \pi y \ .
\ee
We verified all three analytical results with numerical integrations or sums. In  Fig. (\ref{matsubaras}) we plot
the result of Eq. (\ref{matsubara}) and it is clear that in the vanishing $T$ limit all
three analytical results coincide.

\subsection{Infrared regularization of the linear confining potential}

\begin{figure}[t!]
\begin{center}
\includegraphics[width=0.9\columnwidth]{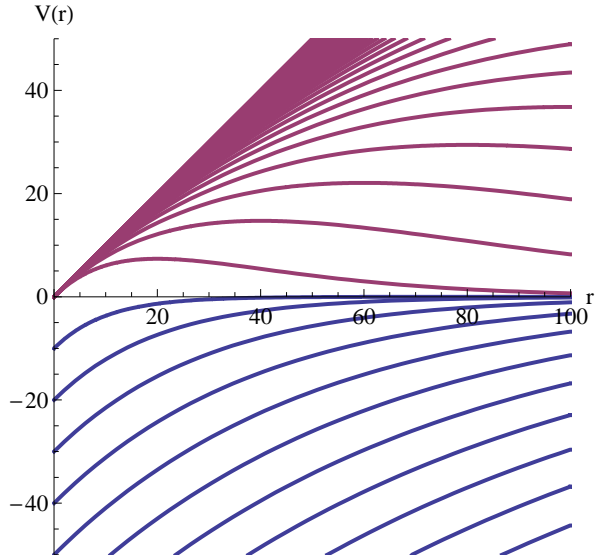}
\caption{The linear potential must be regularized for the Fourier Transform,
and we show our two different regularizations of Eqs. (\ref{IRdivpot}) and (\ref{IRfinpot}),
both leading to $V(r) \to \sigma r $ plus a constant when $r \to 0$.
Two successions of curves are plotted, to illustrate the limit of a vanishing infrared regulator $\lambda \to 0 $.
The regularization of the negative curves maintains the potential monotonous and $V(\infty)=0$
but adds an infrared negative constant to the potential.
The regularization of the positive curves maintains $V(0)=0$ 
but the potential decreases for large $r$.
\label{linearregu}}
\end{center}
\end{figure}

Notice that in the case of a linear potential, divergent in the
infrared, the Fourier transform needs an infrared
regulator $\lambda$ eventually vanishing.
A possible regularization of the linear potential is,
\be
V(r)= - \sigma {e^{- \lambda \, r} \over \lambda} \simeq - {\sigma \over \lambda} + \sigma r \ ,
\label{IRdivpot}
\ee
corresponding to a model of confinement where the quark-antiquark system has
an infinite binding energy ${\sigma \over \lambda}$at the origin $r=0$, is montonous 
and only vanishes at an arbitrarily large distance. 
This potential has a simple three-dimensional Fourier transform,
\bea
V(k) &=& \int_0^\infty  dr  \, { 4 \pi r \sin( k r ) \over k }  \, V( r)  
\nonumber \\
&=&  \sigma { - 8 \pi \over ( k^2 + \lambda^2)^2   }
\eea
and this is the most common form of the linear potential in momentum space
utilized in the literature. Notice that this is infrared divergent due to the $-1/\lambda$ infinite binding energy
in the limit where the regulator $\lambda \to 0$.
If we want to avoid the infinite binding energy we should use a different
regularization of the linear potential, also vanishing when $r \to \infty$ but
not monotonous since ir grows linearly at the origin starting with $V(0)=0$,
\be
V(r)= \sigma r \, e^{- \lambda \, r} 
\label{IRfinpot}
\ee
where the Fourier transform,
\be
V(k) =\sigma { - 8 \pi \over ( k^2 + \lambda^2)^2   } + \sigma {32 \pi \lambda^2 \over
 ( k^2 + \lambda^2)^3  } 
\label{momIRfinpot}
\ee
is such that the integrals in $k$ no longer diverge. For instance,
\be
\int_{-\infty}^{+ \infty} k^2 dk V(k) = 0
\ee
since this is proportional to $V(0)=0$. The new term in the potential
$ {32 \pi \lambda^2 \over
 ( k^2 + \lambda^2)^3  } $ is equal to $ (2 \pi)^3 \delta^3(k) / \lambda$ in
 the limit $\lambda \to 0$ and thus the potential in Eq. (\ref{momIRfinpot}) is infrared finite.
Both the potentials in Eqs. (\ref{IRdivpot}) and  (\ref{IRfinpot}) are
illustrated in Fig. \ref{linearregu}. In the vanishing temperature limit
$T=0$ the different regularizations lead to the same physical results since
any constant term in a density-density interaction has no effect in the quark 
running mass $m(p)$ or in the hadron spectrum
\cite{Bicudo:2010qp}.

With the potential of Eq. (\ref{IRdivpot})
the functions $I_B$ and $I_A$ contributing to the mass gap
equation and to the one quark energy are,
\bea
I_A(k,p,\lambda) &=& { p k \over (p-k)^2 + \lambda^2} 
- { p k \over (p+k)^2 + \lambda^2} \ ,
\nonumber \\
I_B(k,p,\lambda) &=& 
 { p k \over (p-k)^2 + \lambda^2} 
+ { p k \over (p+k)^2 + \lambda^2}
\nonumber \\ 
&&
+ {1 \over 2} \log  {(p-k)^2 + \lambda^2 \over (p+k)^2 + \lambda^2}\ .
\eea

\subsection{A set of two non-linear equations.}

At finite temperature $T\neq0$ the mass gap equation for the quark running mass $m(p)$ 
couples to the one quark energy equation $E(p)$ through the Matsubara sum, 
and thus we get  a system of three non-linear coupled  equations,
\bea
\label{non-linear}
m_{_T} (p) &=& m_0+ { \sigma(T) \over p^3}
\int_0^\infty {d k \over 2 \pi}  {\cal M}_{_T}(k)
\\
&& {I_A(k,p,\lambda)\, 
 m_{_T} (k) p   -
I_B(k,p,\lambda)\,
 m_{_T} (p) k
\over \sqrt{k^2 + m_{_T} (k)^2}}   \ ,
\non \\
E_{_T}(p) &= &  
{ p^2 + m_0 m_{_T} (p) 
\over   \sqrt{p^2 + m_{_T} (p)^2}  }
+ {\sigma(T) \over p^2}\, \int_0^\infty {d k \over 2 \pi}  {\cal M}_{_T}(k)
\non \\
&& 
{ I_A(k,p,\lambda) 
\,    p \, k   +  I_B(k,p,\lambda) \,   m_{_T} (p) \, m_{_T} (k)  \over   \sqrt{p^2 + m_{_T} (p)^2} \sqrt{k^2 + m_{_T} (k)^2} }\ ,
\non
\eea
where the Matsubara sum normalized to 1 in the limit of small temperatures is,
\bea
{\cal M}_{_T}(p) &=&
2 \sum_{n=- \infty}^{+ \infty} i \, k T
{ i  \over i \, (2 n +1) \pi \, k T - \left[E(\mbf p) - \mu\right]} 
\non \\
&=&   \tanh { E_{_T}(p)  - \mu \over 2 \, k  \, T} \ .
\eea
To solve the coupled systems of non-linear Equations
(\ref{non-linear}), we apply the techniques detailed in 
\cite{Bicudo:2010qp},
{\em i. e.} we utilize a Pad\' e ansatz for the quark running mass $m_{_T}(p)$,
we solve the mass gap variationally, we compute the one quark energy $E_{_T}(p)$,
we fit is with a Pad\' e approximant, we feed it back into the mass gap equation, 
and then we repeat this cycle iteratively until the solution converges. The solution of
the system of Eqs. (\ref{non-linear}) constitutes the main goal of this paper.

\section{The $T\neq0$ solutions of the mass gap equation}

Finite temperature $T$ changes the mass gap equation in two different ways,
in the temperature dependence of the string tension $\sigma(T)$
and in the Matsubara sum replacing the $p^0$ integral of the propagators.
Since the finite temperature mass gap equation is numerically cumbersome to address, 
we first study separately the effect of each of these changes, 
before including both in the mass gap equation.

Note that we work in units of $ \sqrt \sigma = \sqrt{ 0.19}$ GeV, the string tension
commonly used in the Charmonium spectroscopy. 
For the transition temperature, we utilize the result computed in Lattice QCD 
\cite{Aoki:2006we,Aoki:2006br}
of  0.176 (7) GeV. Thus we get $T_c= 0.40 \sqrt \sigma$.

\subsection{The effect of the finite temperature string tension $\sigma(T)$}

Here we consider the $T$ dependence of the string tension only.
Notice this approach is exact when our infrared regulator $U_0 \to \infty$ as in the standard regularization of the linear potential of Eq. (\ref{IRdivpot}). 
In that case $E_{_T}(p) \to \infty$ and the Matsubara sum ${\cal M}_{_T}=1$ is simply the same one occurring at $T=0$ .

\begin{figure}[t!]
\begin{center}
\includegraphics[width=1.\columnwidth]{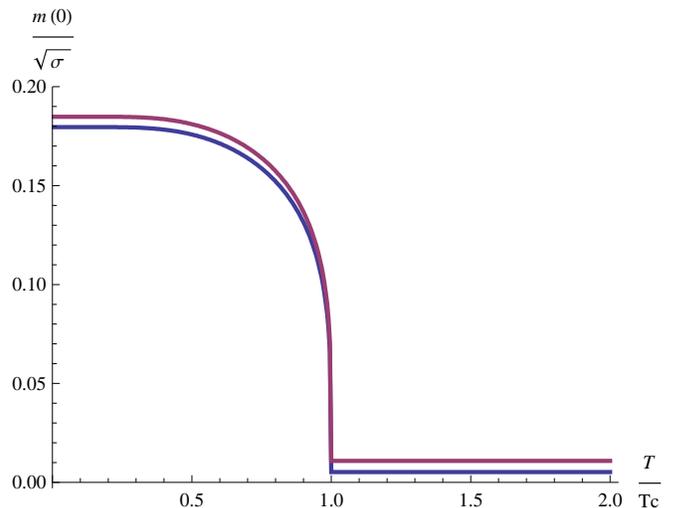}
\caption{Here we consider only the effect of the temperature in the string tension $\sigma(T)$, and plot the mass gap $m(0)$ as a function of $T$ for
the light $u$ (lightest) and $d$ (slightly heavier) quarks. 
We have a crossover, close to a phase transition since the
current masses of the light quarks are much smaller than
the dynamically generated constituent quark mass.
\label{massgapofTlights}}
\end{center}
\end{figure}

Then the finite $T$ mass gap equation is similar to $T=0$ mass gap 
equation, just with a substitution of $\sigma_0$ by $\sigma(T)$. 
The mass gap solution can be found with a simple rescaling 
of the string tension. 

In units of $\sqrt \sigma =1$, we can fit the mass gap at the origin $m(0)$ as
a function of the running mass $m_0$  with a function interpolating from the
chiral limit constant to the massive case linear function with
a simple irrational ansatz,
\be
m(0)= { a + m_0 + \sqrt{ (a -m_0)^2 + 4 b^2} \over 2}
\ee
as depicted in Fig. (\ref{fitmassgap} - letf).
Our fit produces $a= -2.29223 \sqrt \sigma $, $ b= 0.656306 \sqrt \sigma$.
Then our solution is simply found rescaling   $\sqrt \sigma$ from 1 to
$\sqrt{ \sigma(T)}$

\begin{figure}[t!]
\begin{center}
\includegraphics[width=1.\columnwidth]{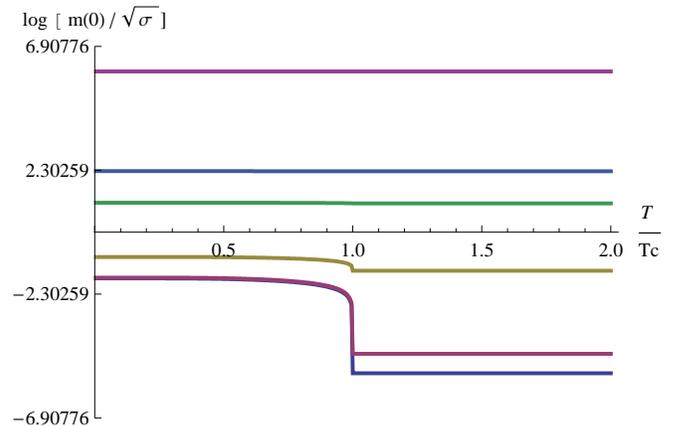}
\caption{Log plot of the mass gap $m_q(0)$, considering $T$ effects in $\sigma(T)$ only, for the six
different flavours of quarks as a function of $T$. From bottom to 
top we show the quarks $u$, $d$, $s$, $c$, $b$ and $t$.
The heavier the quark, the weaker the crossover gets.
\label{massgapofT}}
\end{center}
\end{figure}

Our results, depicted  in Figs. \ref{massgapofTlights} and  \ref{massgapofT},
show a chiral crossover, 
since the mass gap never vanishes  due to the finite current quark mass $m_0$. 
We observe in our results that the heavier the quark, the weaker the chiral crossover gets, as
show in Figs. \ref{massgapofTlights} and  \ref{massgapofT}.
For the heavier quarks the chiral crossover is very weak.
On the other side, of light $u$ and $d$ quark masses, 
we still have a crossover since the
mass gap starts finite at $T=0$ and continues to be finite
beyond $T_c$, although at large $T >> T_c$ the mass gap is 
quite smaller than at vanishing $T$.

\begin{figure}[t!]
\begin{center}
\includegraphics[width=1.\columnwidth]{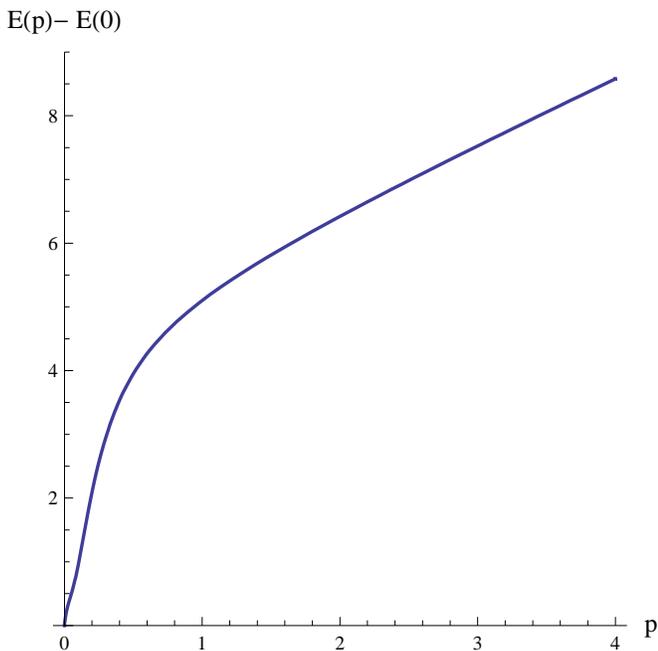}
\caption{A possible scenario to compute the Matsubara sum
consists in using the one quark dispersion relation 
subtracted by the enery at the origin $E(p)-E(0)$ . This produces the smallest
possible quark energy to be used in the Matsubara sum.
The other opposite scenario consists in utilizing the energy $E(p)$,
wint an infrared divergence of $\sigma \over 2 \lambda$.
In any case the one quark energy
\label{onequarkenergy_e0}}
\end{center}
\end{figure}

\begin{figure}[t!]
\begin{center}
\includegraphics[width=1.\columnwidth]{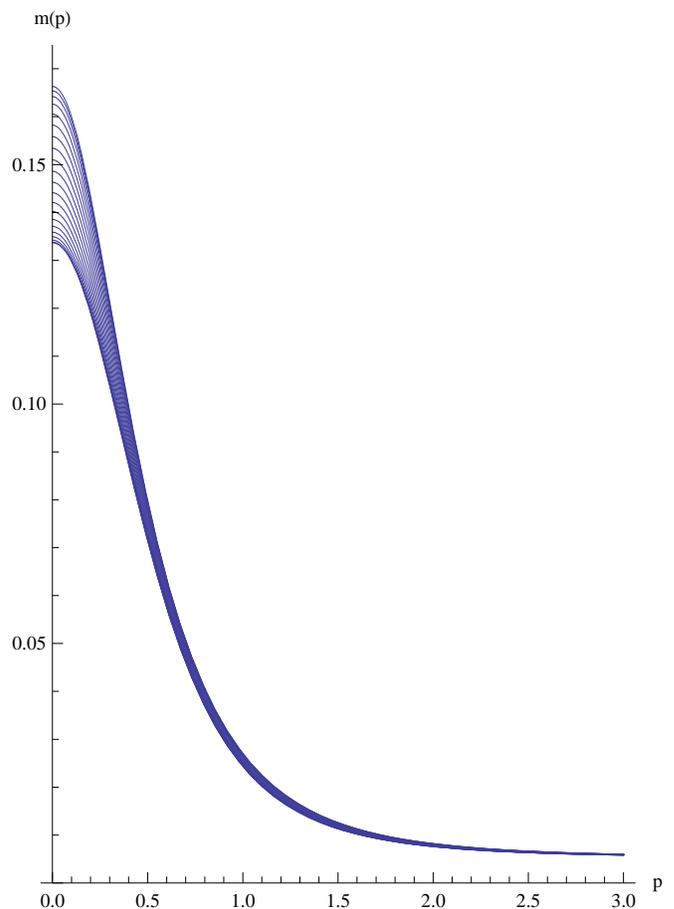}
\caption{Here we ignore the effect of the temperature in the string tension and
consider only the effect of the Matsubara sum ${\cal M}_{_T}$. We plot
the running mass for the $d$ quark , $m_d(p)$,
computed for the different temperatures present in Table   \ref{table 1}, ranging from
$T=0 $ to $T=T_c$.
Note that the temperature decreases the quark mass,
but the effect is much weaker than the string tension effect
in Fig. \ref{massgapofTlights}.
\label{mdofpdifTmtsuonly}}
\end{center}
\end{figure}

Notice that the mass gap $m(0)$ 
changes quite abruptly when $T$ crosses $T_c$. 
The steepness of the light quark mass at $T=T_c$ is due 
to our fit of $\sigma(T)$ which is second-order like,
as in Fig. \ref{F1Kacz}. Possibly with a crossover-like $\sigma(T)$ reflecting 
the dynamical fermion effects, the crossover in the quark mass $m(0)$ 
may get slightly smoother. 
However we do not know exactly 
what happens at the transition temperature for light 
quarks, since an opposite effect can make the string tension steeper again at $T_c$. 
Indeed for light quarks one may argue that the internal energy $U_1$ (steeper than 
the free energy $F_1$) should be used. Thus in this paper we cannot address 
in detail the exact temperature where $T$ crosses $T_c$, but rather the remaining
range of temperatures.

\subsection{The effect of the Matsubara sum effect in the propagator}

Here we are interested in finding the effect of the Matsubara sum in the mass gap equation.
To isolate this effect, we neglect any effect of the temperature $T$ on the string tension,
considering the string tension $\sigma_0$ at all temperatures. 

We must then choose an infrared regulator, since adding a constant term $-U_0$ to the potential, 
say $- \sigma / \lambda$ as in Eq. (\ref{IRdivpot}), affects the energy  which is then shifted $E_{_T}(p) \to E_{_T}(p)  + U_0 /2$.
Thus at finite $T$, the solution of the system of Eqs.  (\ref{non-linear})
depends on a constant shift of the potential, in contradistinction  with vanishing temperature where 
${\cal M}_{_T}(p) \simeq 1$ decouples the energy $E_{_T}(p)$ from the mass
\cite{Bicudo:2010qp}. 
At finite $T$ the one quark energy $E_{_T}(p)$ contributes non-linearly to the Matsubara sum $D_{_T}(p)$, 
and any change in  $E_{_T}(p)$ does change the solution of the mass gap and energy coupled equations and thus the three Eqs.  (\ref{non-linear}) are coupled. 

Here we choose a minimal $U_0$, closer to the regularization 
of the linear potential of Eq. (\ref{IRfinpot}), just sufficient  to cancel the one quark energy
at vanishing momentum $E_{_T}(0)=0$. This case is also equivalent to the one when a chemical 
potential $\mu=E_{_T}(0)$ is included, and thus $E_{_T}(0) - \mu=0$.
We consider this case since it produces the largest possible effect in the Matsubara sum $D_T(p)$, interpolating between $0$ at vanishing momentum and $1$ at large momentum.   
This does suppress the infrared part of the integral if the mass gap equation up to $E_{_T}(k) -E_{_T}(0)  \simeq kT$.

To exactly cancel the infrared divergences in the integrals, we utilize ansatze for the quark mass
$m_{_T}(p)$ and for the quark energy $E_{_T}(p)$. Our numerical technique was tested
in great detail for $T=0$ an here we utilize the same ansatze and numerical techniques
\cite{Bicudo:2010qp}.
Our ansatze for $m_{_T}(p)$ and $E_{_T}(p)$ are the Pad\'e approximants,
\bea
m_{_T}(p ) &=& {1 \over c_0 + c_2 p^2 + c_4 p^4} \ ,
\non \\
E_{_T}(p) & = & {\sigma \over 2 \lambda} - { 2 \sigma \over \pi} { p \over  p^2 - m_{_T}(p)^2 } 
\non \\
&& + p C(p) + m_0 S(p)  + \widetilde E_{_T}(p)  \ ,
\non \\
\widetilde E_{_T}(p) &=& - {1 \over e_0 +e_1 p + e_3 p^3 + e_5 p^5} \ .
\non 
\label{ansatze}
\eea
After each iteration the resulting $m_{_T}(p )$ and $E_{_T}$ are fitted again
with our Pad\'e approximants, and a new iteration is started.
But for teh solution for each temperature and mass is completely determined by the three parameters for the fits of 
$m_{_T}(p )$,  since $E_{_T}(p )$ is a function of the quark mass.

\begin{figure}[t!]
\begin{center}
\includegraphics[width=1.\columnwidth]{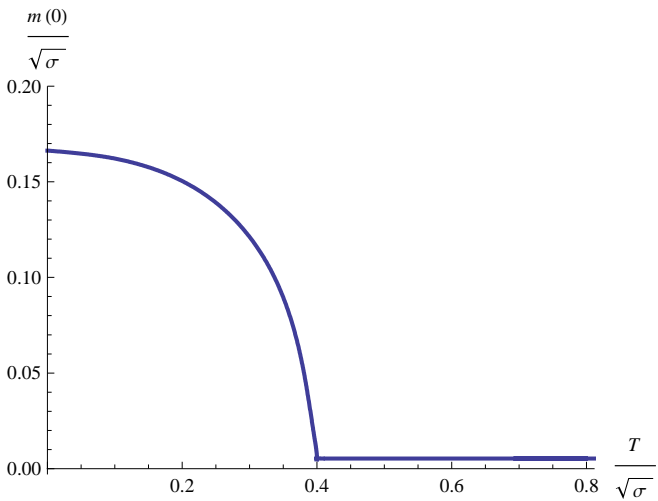}
\caption{Full computation of the mass gap for the $d$ quark , $m_d(0)$, 
in the infrared regulator case where the Matsubara sum effects are maximal. 
Although not as steep close to $T=T_c$, the curve is quite similar to the one where we ignore the Matsubara sum effects. 
Again we have a crossover at $T_c$.
\label{dmassgapfull}}
\end{center}
\end{figure}

To ensure convergence,
we utilize the technique 
\cite{Bicudo:2010qp}
of translating the mass gap equation into the variational equation for 
the vacuum energy density,
\bea
\label{densityofT}
 {\cal E} &=& { -g \over 2 \pi}\int_0^\infty  {dp \over 2 \pi} D_{_T}(p) 
 \biggl\{
2p^3 C(p) + 2 p^2 m_0 S(p)  
\non \\ 
&& + \sigma(T) \
\int_0^\infty {d k \over 2 \pi} D_{_T}(k) 
  \biggl[ I_A(k,p,\mu) \,   S(k) S(p) 
\non \\ 
&&
\hspace{2cm}
+ I_B(k,p,\mu) \,   C(p) C(k) \biggr]  \biggr\}\ ,
\eea
now with a new dependence on temperature in $\sigma_T$ (momentarily we consider $\sigma_T=\sigma_0$)
and in the Matsubara sum ${\cal M}_{_T}$ .
When the current mass is small compared with the typical scale $\sqrt{ \sigma(T)}$ 
of our problem, we utilize
directly this equation. In the opposite case were the string tension is much smaller
than the current mass, close to $T_c$, then the solution $m(p)$ is close to the current mass $m_0$,
\be
m(p) \simeq m_0 + {\sigma \over m_0} {\cal F} ({p \over m_0})  \ ,
\ee
and then we only need to compute the dimensionless function ${\cal F}(p)$,
produced pertubatively
\cite{Bicudo:2010qp}
after few iterations of the fixed point equation.

We consider the case of $m_0 = m_d$ corresponding to a $d$ quark.
Our results are illustrated in Fig. \ref{mdofpdifTmtsuonly}.
Note that the effect of the Matsubara sum ${\cal M}_{_T}$ is to decrease the quark
mass with increasing temperature, 
but nevertheless the  Matsubara sum effect in Fig. \ref{mdofpdifTmtsuonly} is much weaker than the string tension effect shown in Fig. \ref{massgapofTlights}.

\begin{figure}[t!]
\begin{center}
\includegraphics[width=1.\columnwidth]{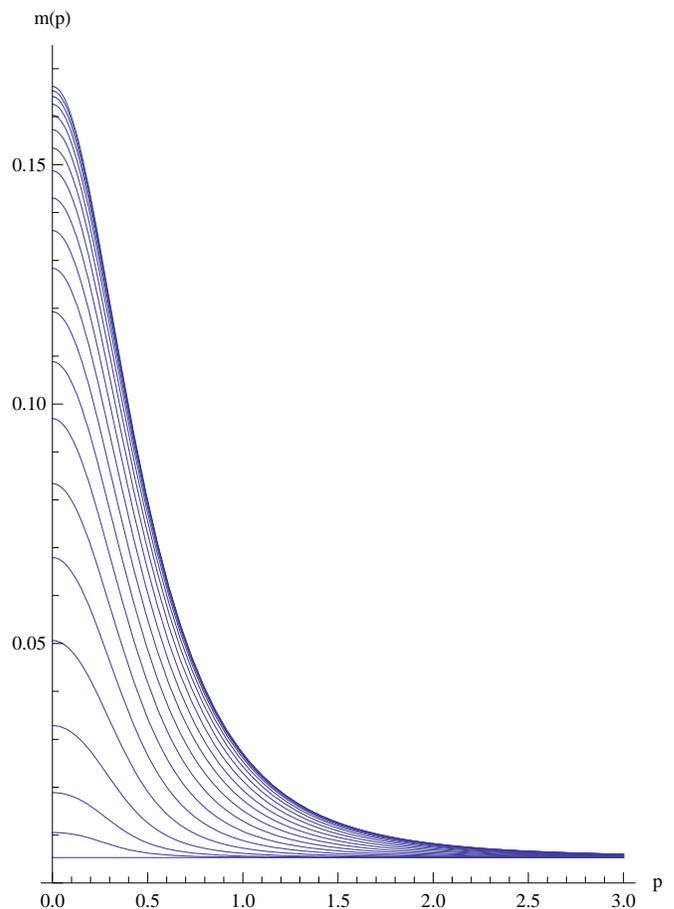}
\caption{Full computation of the running mass for the $d$ quark , $m_d(p)$,
computed for the different temperatures present in Table   \ref{table 1}, ranging from
$T=0 $ to $T=T_c$.
Unlike when the Matsubara sum only is considered, here the quark mass indeed equals the
current quark mass $M_d$, almost vanishing, for $T=T_c$.
\label{mdofpdifTs}}
\end{center}
\end{figure}

\subsection{Full solution of the mass gap equation at finite temperature $T$}

Here we determine the full solution of the mass gap equation at finite temperature $T$,
utilizing Eq. (\ref{densityofT}), but combining the temperature effects in the string tension $\sigma(T)$ 
with the temperature effects in the Matsubara sum ${\cal M}_{_T}(p)$.
In the present case we continue to consider that $\mu=E(0)$, the case maximizing the finite temperature effects of the
Matsubara sum.

We now fix the current quark mass to $m_0 = m_d$, quark $d$ current mass 
and run the temperature $T$ so that we get a curve.
We numerically consider a set of temperatures denser at $T_c$, with the parametrization of $T=T_c \cos{\theta}$ 
where the $\theta$ considered are equally spaced in the interval ranging between $\pi/2$ and $0$.
We show the best fitting parameters $c_0, c_2, c_4,$ in Table \ref{table 1}, 
we depict the mass gap critical curve in Fig. \ref{dmassgapfull}
and we show the different running quark masses in Fig. \ref{mdofpdifTs} .
It occurs that indeed a crossover is found in the mass gap critical curve at $T=T_c$, 
and that the critical curve is similar (but not identical) to the critical curve computed with the string tension temperature effects only.

\section{Conclusion}

\begin{table}[t!]
\begin{center}
\begin{tabular}{c c | c c c c}
\hline 
$m_0$ & $T$ & $ 10^4 {{\cal E} - {\cal E}_0 \over g} $ & $c_0$ & $c_2$ & $c_4$ 
 \\ \hline 
0.00527656  &  0.000000  &  -0.558232  &  6.20966  &  25.2077 
 &  14.5339 \\
   0.00527656  &  0.0313837  &  -0.514877  &  6.24603  &  25.1480  &  
14.6870 \\
   0.00527656  &  0.0625738  &  -0.482888  &  6.29268  &  25.0808  &  
14.8787 \\
   0.00527656  &  0.0933782  &  -0.456189  &  6.35722  &  25.0130  &  
15.1468 \\
   0.00527656  &  0.123607  &  -0.425455  &  6.45135  &  25.0128  &  
15.5903 \\
   0.00527656  &  0.153073  &  -0.390516  &  6.57955  &  25.2058  &  
16.3132 \\
   0.00527656  &  0.181596  &  -0.35179  &  6.74966  &  25.6934  &  
17.4403 \\
   0.00527656  &  0.208999  &  -0.309921  &  6.97245  &  26.5531  &  
19.1262 \\
   0.00527656  &  0.235114  &  -0.265958  &  7.26188  &  27.8649  &  
21.5887 \\
   0.00527656  &  0.259779  &  -0.221304  &  7.63714  &  29.7358  &  
25.1689 \\
   0.00527656  &  0.282843  &  -0.177546  &  8.12658  &  32.3267  &  
30.4365 \\
  0.00527656  &  0.304162  &  -0.136292  &  8.77490  &  35.8925  &  
38.3987 \\
 0.00527656  &  0.323607  &  -0.0990367  &  9.65716  &  40.8552  &  
50.9525 \\
  0.00527656  &  0.341056  &  -0.0670518  &  10.9094  &  47.9602  &  
71.9645 \\
  0.00527656  &  0.356403  &  -0.0412995  &  12.8039  &  58.6492  &  
110.150 \\
  0.00527656  &  0.369552  &  -0.0223216  &  15.9619  &  76.1081  &  
187.827 \\
  0.00527656  &  0.380423  &  -0.0100832  &  22.0570  &  108.769  &  
371.573 \\
  0.00527656  &  0.388948  &  -0.00365887  &  36.2780  &  183.098  &  
886.265 \\
  0.00527656  &  0.395075  &  -0.00105524  &  73.6460  &  382.799  &  
2491.37 \\
  0.00527656  &  0.398767  &  -0.000187953  &  190.883  &  1073.65 
 &  9021.24 \\
 0.00527656 & 0.400000 & -0.0000000 &
$\infty$ &
$\infty$ &
$\infty$
\\ \hline
\end{tabular}
\caption{
The parameters for the quark functions $m_{_T}(p )$
defined in Eq. (\ref{ansatze}).
\label{table 1} }
\end{center}
\end{table}

We apply a new variational technique, in the framework of a \q,  to solve the mass gap equation at finite temperature $T$. 
The quark mass $m_{_T}(p)$ and the quark energy $E_{_T}(p)$ are fitted with Padé approximants, the quark mas parameters are displayed in Table \ref{table 1}. 

We compare the two different temperature contributions, in the string tension $\sigma(T)$ and in the Matsubara sum ${\cal M}_{_T}(p)$. It occurs that the dominant contribution, and also the simplest one to apply in the quark model, 
is the one of the string tension $\sigma(T)$.  This happens not only because the deconfinement critical temperature $T_c = 0.40 \sqrt \sigma$ is relatively small 
when compared to the string tension at vanishing temperature $\sigma_0$, 
but also because the Matsubara  sum never vanishes while the string tension really vanishes when $T \to T_c$.
Thus the quark mass critical curve has a shape similar to the string tension critical curve, 
but the curves are not exactly identical since the quark critical curve is less steep at $T \simeq T_c$.

Moreover, since the light current quark masses $m_u$ and $m_d$ are small compared with the string tension at vanishing temperature $\sigma_0$, the quark mass gap $m(0)$ essentially follows the string tension curve. This is a quite simple result, relevant for further studies at finite temperature.

With our excellent fits of  the dynamical quark mass $m_{_T}(p)$ and of the quark dispersion relation $E_{_T}(p)$  we are now well equipped to address further studies, 
such as the hadronic excited spectra at finite temperature $T$.

 \hspace{1cm}
\begin{acknowledgments}
I am very grateful to Marlene Nahrgang, to Pedro Sacramento and to Jan Pawlowski 
for dicussions on the QCD phase diagram motivating this paper. 
I acknowledge the financial support
of the FCT grants CFTP, CERN/FP/109327/2009 and
CERN/FP/109307/2009.
\end{acknowledgments}


\end{document}